\documentclass[%
preprint,aps,
nofootinbib,amsmath,amssymb,amsfonts,
superscriptaddress
]{revtex4-2}
\usepackage{graphicx}
\usepackage{bm}
\usepackage{xcolor}
\usepackage{comment}
\usepackage{caption}
\usepackage{subcaption}
\usepackage{hyperref}
\usepackage{float}
\usepackage[section]{placeins}
\hypersetup{
    colorlinks=true,       
    linkcolor=red,          
    citecolor=blue,        
}
\begin{document}
\title{Non-trivial quantum fluctuations in asymptotically non-flat black-hole space-times}
\author{Manu Srivastava}
\email{manu$_$s@iitb.ac.in}
\affiliation{Department of Physics, Indian Institute of Technology Bombay, Mumbai 400076, India} 
\author{S. Shankaranarayanan} \email{shanki@phy.iitb.ac.in}
\affiliation{Department of Physics, Indian Institute of Technology Bombay, Mumbai 400076, India}
\date{\today}
\begin{abstract}
Hawking radiation remains a crucial theoretical prediction of semi-classical gravity and is considered one of the critical tests for a model of quantum gravity. However, Hawking’s original derivation used quantum field theory on a fixed background. Efforts have been made to include the space-time fluctuations arising from the quantization of the dynamical degrees of freedom of gravity itself and study the effects on the Hawking particles. Using semi-classical analysis, we study the effects of quantum fluctuations of scalar field stress-tensors in asymptotic non-flat spherically symmetric black-hole space-times. Using two different approaches, we obtain a critical length-scale from the horizon at which gravitational interactions become large, i.e., when the back reaction to the metric due to the scalar field becomes significant. For 4-D Schwarzschild AdS (SAdS) and Schwarzschild de Sitter (SdS), the number of relevant modes for the back-reaction is finite \emph{only} for a specific range of values of $M/L $ (where $M$ is the mass of the black-hole, and $L$ is related to the modulus of the cosmological constant). For SAdS (SdS), the number of relevant modes is infinite for 
$M/L \sim 1$ ($0.2 < M/L < 1/(3 \sqrt{3})$. We discuss the implications of these results for the late stages of black-hole evaporation.
\end{abstract}
\maketitle

\newpage
\section{Introduction}

Hawking radiation is one of the most striking effects widely agreed to arise from quantum theory and general relativity~\cite{1974-Hawking-Nature,1975-Hawking-CMP}. Hawking radiation arises from quantizing matter fields in a fixed background space-time with an event horizon. In other words, black-holes constantly radiate particles with energies of the order of Hawking temperature (which is proportional to the surface gravity of the horizon)~\cite{1976-Davies-PRS,1993-Page-PRLa,1994-Bekenstein-Arx}. 
Hawking's original derivation makes two key assumptions: First, the radiation which emerges at late times is strongly redshifted as it climbs out of the black-hole's gravitational well. In other words, 
the derivation 
requires incoming vacuum modes with frequencies far above the Planck scale~\cite{1991-Jacobson-PRD,1993-Jacobson-PRD,1995-Brout.etal-PRDa,1995-Unruh-PRD}. Second, the derivation ignores the interactions of the radiation with matter~\cite{1992-Susskind.Thorlacius-NPB}.

In the modern viewpoint, general relativity is an effective description valid below a particular cutoff momentum scale~\cite{1994-Donoghue-PRD,1995-Donoghue-Proc,2004-Burgess-LRR}. In the effective field theory picture, an effective Lagrangian (valid below some scale $E < \Lambda$) 
\begin{equation}
\label{eq:EFT}
-\frac{\mathcal{L}_{\mathrm{eff}}}{\sqrt{-g}}=
\frac{M_{\rm Pl}^{2}}{2} R+a R_{\mu \nu} R^{\mu \nu}+b R^{2}+d R_{\mu \nu \lambda \rho} R^{\mu \nu \lambda \rho}+e \square R+\frac{c}{\Lambda^{2}} R^{3}+\cdots
\end{equation}
accounts for all tree and loop level corrections from particles of masses greater or equal to $\Lambda$~\cite{2020-Ruhdorfer.etal-JHEP}. $a, b, c, d,$ and $e$ are dimensionless constants and $M_{\rm Pl}$ is the Planck mass. However, these do not include the gravitational effects generated from quantum loops of known particles. In principle, these can be computed and are known to be finite and free from renormalization ambiguities~\cite{1984-Birrell.Davies-Book}.  While these effects are, of course, highly suppressed in space-times without horizons, it has been shown that these corrections can be large close to the horizons~\cite{Casher,1999-Sorkin.Sudarsky-CQG,2000-Barrabes.etal-PRD,2007-Giddings-PRD,2007-Hu.Roura-PRD,2008-Thompson.Ford-PRD}.

Recently, Almheiri et al. \cite{2013-Almheiri.etal-JHEP} used a similar argument and proposed a firewall near the horizon to resolve the inconsistency in black-hole complementarity~\cite{1992-Susskind.Thorlacius-NPB}. At infinity, Hawking radiation is dominated by low angular momentum modes, since high angular momentum modes are trapped behind a potential barrier. More specifically, they showed that an infalling {particle} would encounter a divergent stress tensor at the stretched horizon~\cite{1992-Susskind.Thorlacius-NPB} and, hence, the need for a firewall. 
However, in these calculations, it is assumed that the matter fluctuations do not significantly alter the outgoing modes that carry the thermal radiation to distant observers. 

In an attempt to address this issue, in this work, we evaluate the quantum fluctuations of the massless scalar field in asymptotically non-flat black-hole space-times. In comparison, the effect of quantum fluctuations of the massless scalar field in an asymptotically flat black-hole space-times has been performed earlier~\cite{Casher,2008-Thompson.Ford-PRD}, such an analysis has not been performed for asymptotic AdS or dS space-times. In this work, we obtain a length scale from the horizon at which gravitational interactions become large, i.e., when the back reaction to the metric due to the scalar field becomes significant. Within this length scale, Hawking's assumption~\cite{1974-Hawking-Nature,1975-Hawking-CMP} of a free field propagating on a given classical background breaks down. In the 4-D Schwarzschild AdS case, we show that the semiclassical analysis to account for the back-reaction breaks down in the limit $M/L \sim 1$ where $M$ is the mass of the black-hole and $L$ is related to the modulus of the cosmological constant. [This is the limit considered in AdS/CFT correspondence where the large black holes in
AdS are dual to the high-temperature phase of the dual field theory~\cite{Hubeny}.] In the 4-D Schwarzschild-de Sitter case, we show the analysis can only be employed for the event horizon and breaks down in the Nariai limit~\cite{Shankaranarayanan:2003ya}.

We calculate the back-reaction of the fluctuations on the horizon using two different approaches. In the first approach, we use the statistical mechanical properties of the perturbations and calculate the change in the mass of the black-hole due to these fluctuations. In the second approach, we calculate the effect of a shockwave (due to an infalling massless particle) on the outgoing Hawking particle~\cite{Shockwave, Aichelburg:1970dh}. 
While the two approaches use two different features of the fluctuations, we show that they provide similar results when the semiclassical analysis is valid. We discuss the limitation of our approach near the cosmological horizon in the case of Schwarzschild de Sitter (SdS) and when $M \sim L$ for Schwarzschild Anti-de Sitter (SAdS).

In Sec.~\eqref{sec:II}, we discuss the kinematical properties of SAdS and SdS. In Sec.~\eqref{sec:III}, we introduce the setup that we use to calculate the back-reaction of the quantum fluctuations on a 4-D spherically symmetric black-hole space-time. In Secs.~\eqref{sec:IV} and \eqref{sec:V}, we obtain the back-reaction using two different approaches. In Sec. \eqref{sec:Limitations}, we discuss the limitations of the analysis in $M/L \sim 1$ limit. The two Appendices contain the details of the calculations of Approach 2. We use $(-, +, +, +)$ signature for the 4-D space-time metric~\cite{1973-Misner.etal-Gravitation}. We use the geometric units $ G = c =1$, and to match the dimensions, we retain the Boltzmann constant ($k_B$) and Planck's constant ($\hbar$).

\section{Kinematical properties of the black-hole space-times}
\label{sec:II}

We consider the following general 4-dimensional spherically symmetric metric:
\begin{equation}
\label{eqn:SSmetric}
    d s^{2}=-f(r) d t^{2}+\frac{d r^{2}}{f(r)}+r^{2} d \Omega^{2}
\end{equation}
with $d \Omega^{2}$ being the usual $2$-sphere element. 
In this work, we consider asymptotically non-flat space-times. Specifically, we consider two cases: Schwarzschild Anti-de Sitter (SAdS) and Schwarzschild de Sitter (SdS). 
In the former we have~\cite{Hubeny}:
\begin{equation}
\label{eqn:SAdS_f(r)}
    f(r)=1-\frac{2M}{r}+\frac{r^2}{L^2}
    =1+\frac{r^2}{L^2}-\frac{r_h}{r}\left(\frac{r_h^2}{L^2}+1\right)
\end{equation}
For SdS, we have~\cite{Shankaranarayanan:2003ya}:
\begin{equation}
\label{eqn:SdS_f(r)}
    f(r)=1-\frac{2M}{r}-\frac{r^2}{L^2}
\end{equation}
where $M$ is the mass of the black-hole and $\pm L^{-2}$ corresponds to the measure of curvature in the asymptotic limit of
$dS_4$ and $AdS_4$. In the limit of $L \to \infty$, the above two line-elements reduce to Schwarzschild black-hole (which is a spherically symmetric black-hole in asymptotically flat space-time).  

Using the property that the event horizon is a null hypersurface, the horizon ($r_h)$ is determined by the condition $g^{\mu\nu} \, \partial_{\mu} N \, \partial_{\nu}N = 0$.  For the line-element (\ref{eqn:SSmetric}), $N$ is a function of $r$ and $g^{\mu\nu} \, \partial_{\mu} N \, \partial_{\nu}N = 0$ leads to $f(r) = 0$.  Thus, the location of the event horizon is given by the roots of the equation $f(r)=0$.
SAdS line-element \eqref{eqn:SAdS_f(r)} supports only one real positive root and is given by 
\begin{equation}
    \chi =\frac{r_h}{L}=\frac{2}{\sqrt{3}}  \sinh \left[\frac{1}{3} \sinh ^{-1}\left(3 \sqrt{3} \frac{M}{L}\right)\right]
\end{equation}
In the limit of $L \to \infty$, the above event horizon 
reduces to $r_h=2M$. The horizon radius is denoted as $r_h$. 

For $0 < \frac{M}{L}<\frac{1}{3\sqrt{3}}$, SdS line-element \eqref{eqn:SdS_f(r)} supports two real positive zeroes~\cite{Shankaranarayanan:2003ya}, and are given by:
\begin{equation}
    \label{eqn:chi_dS}
    \chi_b=\frac{r_b}{L}=\frac{2}{\sqrt{3}} \cos\left(\frac{\pi +\psi}{3}\right)\quad\quad\quad \chi_c=\frac{r_c}{L}=\frac{2}{\sqrt{3}} \cos\left(\frac{\pi -\psi}{3}\right) \, ,
\end{equation}
where 
\begin{equation}
    \psi= \cos^{-1}\left(3\sqrt{3}\frac{M}{L}\right)
\end{equation}
$r_b$ corresponds to the event horizon of the black-hole, 
while $r_c$ refers to the cosmological horizon. In the range 
$0 < \frac{M}{L} <\frac{1}{3\sqrt{3}}$, $\chi_b$ takes the values in (0,$\frac{1}{\sqrt{3}}$). 

Since we want to obtain the length-scale in which the back-reaction of the fluctuations on the horizon can not be ignored, evaluating the quantities in terms of $\chi, \chi_c, \chi_b$ (instead of other parameters like $M/L$) enables exact calculations. In the next section, we will discuss this aspect.

To compare and contrast our results with asymptotically flat space-time black-holes, we define the ratios of quantities for the SAdS (SdS) black-holes to the quantities for the Schwarzschild black-hole. Physically, this corresponds to the two black-holes with identical mass. We denote these quantities with a subscript $R$ (for the ratio) and a superscript SAdS (SdS). For instance, the dimensionless surface gravity for
SAdS is given by:
\begin{equation}
    \kappa^{SAdS}_R=\frac{\kappa^{SAdS}}{\kappa^S}=(1+\chi^2)(3\chi^2+1)
\end{equation}
where the surface gravity for the black-hole line-element \eqref{eqn:SSmetric} is given by:
\begin{equation}
\label{eqn:kappaGeneral}
    \kappa=\frac{1}{2}\left(\frac{d f(r)}{d r}\right)_{r=r_{h}}
\end{equation}
and the corresponding Hawking temperature $T$ is 
\begin{equation}
\label{def:HawkingTemp}
    T=\frac{\hbar \kappa}{2\pi k_B}
\end{equation}

\begin{figure}[!htb]
\begin{subfigure}{.5\textwidth}
  \includegraphics[width=1\linewidth]{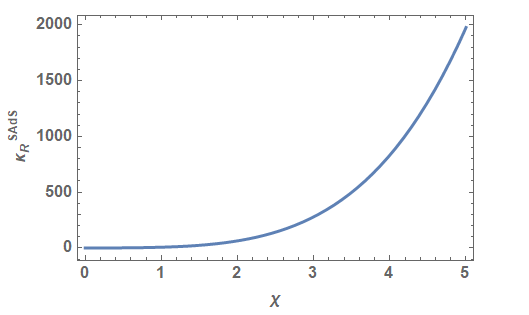}
  \caption{SAdS }
  \label{fig:SurfaceGravityRatioSAdS}
\end{subfigure}%
\begin{subfigure}{.5\textwidth}
  \includegraphics[width=1\linewidth]{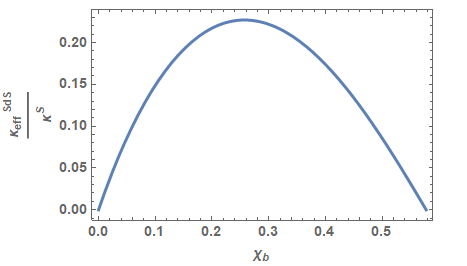}
  \caption{SdS}
  \label{fig:SurfaceGravityRatioSdS}
\end{subfigure}
\caption{Plot of ration of surface gravity with $\chi$ for SAdS and SdS.}
\label{fig:SurfaceGravityRatio}
\end{figure}

For the SdS case, there exists 2 horizons and the effective surface gravity is given by \cite{Shankaranarayanan:2003ya}:
\begin{equation}
    \label{eqn:effKappa_dS}
    \kappa_{eff}^{-1}=|\kappa^{-1}_b|+|\kappa^{-1}_c|
\end{equation}
where $\kappa_b$ and $\kappa_c$ are calculated using \eqref{eqn:kappaGeneral} at the black-hole horizon and the cosmological horizon, respectively.
In Fig \ref{fig:SurfaceGravityRatio}, we have plotted $\kappa^{SAdS}_R$ and $\kappa^{SdS}_{\rm eff}$ as a functions of $\chi$. Note that as $\chi$ increases, the black-hole horizon size increases.

Our aim is to obtain a length scale at which the back-reaction due to the quantum scalar field becomes significant. In order to do that we need to obtain a line-element near the horizon for any spherically symmetric space-time. To do that, we make the 
coordinate transformation $(t, r)~\to~(t,\gamma)$, given by:
\begin{equation}
\label{eqn:gammaDef}
\gamma \equiv \frac{1}{\kappa} \sqrt{f},\qquad d\gamma = 
\frac{1}{2\kappa} \frac{d_r f}{\sqrt{f}} \, dr \, ,
\end{equation}
where $\kappa$ is given by (\ref{eqn:kappaGeneral}). In the above local transformation, the horizon ($r_h$) is at 
$\gamma = 0$. The line-element \eqref{eqn:SSmetric} 
becomes~\cite{2004-Das.Shanki-CQG}:
\begin{equation}
ds^2 = - \kappa^2 \gamma^2 dt^2 + 
4 \,\frac{\kappa^2}{(d_r f)^2}
d\gamma^2 + r^2 d\Omega^2 \, ,
\end{equation}
and hence, near the horizon, we have
\begin{equation}
\label{eqn:nearHorizonMetric}
ds^2 \to - \kappa^2 \gamma^2 dt^2 + d\gamma^2 
+ r_0^2 \, d\Omega^2 \, .
\end{equation}
For space-times with non-degenerate horizons (like SAdS and SdS for which $\kappa \neq 0$), $f(r)$ can be expanded around $r_h$ as $f(r) = f'(r_h) ~ (r - r_h)$. 
%
%
Near the horizon, the invariant distance \eqref{eqn:gammaDef} becomes:
    \begin{equation}
    \label{eqn:GammaDeltaRel1}
        \gamma\sim\frac{2}{\sqrt{f'(r_h)}}\sqrt{r-r_h}=\left(\frac{2(r-r_h)}{\kappa}\right)^{1/2}
    \end{equation}
The line-element \eqref{eqn:nearHorizonMetric} describes the near horizon geometry for a generic spherically symmetric black-hole space-time with any asymptotic structure. This relation is crucial for one of the approaches to obtain the critical invariant distance~\cite{Casher}. 
We will discuss more on this in Sec. \eqref{sec:V}.

\section{Setup}
\label{sec:III}

In this section, we provide the setup to evaluate the invariant distance near the horizon where the back-reaction due to the quantum field becomes significant. The quantum field we consider is a 
massless, minimally coupled scalar field propagating in the spherically symmetric background \eqref{eqn:SSmetric}. The evolution equation of the scalar field is given by: 
\begin{equation}
\label{eq:KGequation}
    \square \Phi=\frac{1}{\sqrt{-g}} \partial_{\mu}\left(\sqrt{-g} g^{\mu \nu} \partial_{\nu} \Phi\right)=0
\end{equation}
The rotational symmetry of the line-element \eqref{eqn:SSmetric},
allows us to decompose into the normal modes $u_{l}(r, \omega)$ (where,
$\omega = E/\hbar$) of 
the field $\Phi$ as follows:
  \begin{equation}
      \phi_{l m}=\frac{u_{l}(r, \omega)}{r} Y_{l m}(\theta, \phi) e^{-i \omega t} 
  \end{equation}
  Putting this back in the wave equation we get:
  \begin{equation}
  \label{eqn:SchLikeEq}
     \left( \frac{d^{2}}{d r_{*}^{2}}  + \omega^2  -V(r) \right) u_{l}(r, \omega)=0
  \end{equation}
where 
\begin{equation}
r_* = \int \frac{dr }{f(r)}
\label{eq:rela-xr}
\end{equation}
denotes the tortoise coordinate, and the effective 1-D potential (often referred to as Regge-Wheeler potential) $V(r)$ is:
\begin{equation}
\label{Eqn:RGPotential}
    V(r)=f(r)\left [ \frac{1}{r} \, \frac{d f(r)}{dr} +\frac{l(l+1)}{r^2} \right ]
\end{equation}

\begin{figure}
\begin{subfigure}{.50\textwidth}
  \includegraphics[width=1\linewidth]{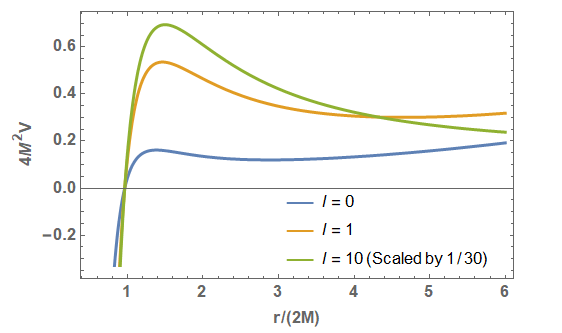}
  \caption{SAdS }
  \label{fig:rhocrit}
\end{subfigure}%
\begin{subfigure}{.50\textwidth}
  \includegraphics[width=1\linewidth]{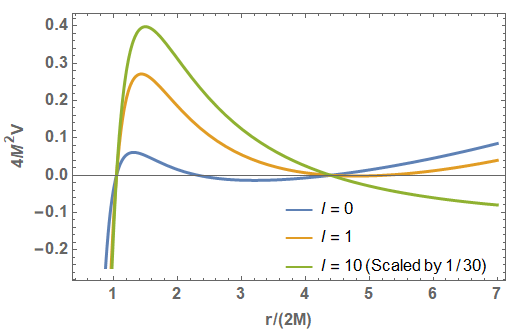}
  \caption{SdS}
  \label{fig:sub2}
\end{subfigure}
\caption{Plot of effective 1-D potential $(4 M^2 V)$ for SAdS and SdS black-holes (${M}/{L}=0.1$).}
\label{fig:PotentialSAdS_SdS}
\end{figure}
The potential $V(r)$ has maxima near event horizon for SAdS and SdS cases and can be treated as an effective 1-D quantum mechanical problem of a particle in a potential barrier. Fig.~\ref{fig:PotentialSAdS_SdS} is the plot of dimensionless potential $(4 M^2 V)$ with dimensionless radial coordinate $(r/(2M))$. We have set ${M}/{L}=0.1$. For both the cases, near the event horizon, the barrier height increases with increasing $l$. Thus, large $l$ modes get trapped near the horizon.

Our analysis rests upon the fact that there is a peak (positive barrier) in the potential profile that traps the high angular momentum modes close to the horizon. For SAdS black-hole, there is no maximum for $\chi \sim 1$ 
(also $M/L \sim 1$). In SdS, there is no maximum near the cosmological horizon $(r_c)$ for any value of $M/L $. We relegate the implication of this condition to  Sec. \ref{sec:Limitations}. We assume that the effective 1-D potential $V(r)$ has a barrier for the rest of the analysis.

The modes that contribute to the energy or mass content of the black-hole (as seen from infinity) are the scattering modes of \eqref{eqn:SchLikeEq}. The condition for such modes is:
\begin{equation}
\label{eqn:OmegaGreaterPotential}
    \omega^2>V(r)
\end{equation}
As mentioned earlier, the black-holes constantly radiate particles with energies of the order of Hawking temperature. Hence, we take $\omega \sim \kappa$. Using the relation \eqref{eqn:kappaGeneral}, we have:
\begin{equation}
  \left(\frac{f'(r_{h})}{2}\right)^2>f(r)\left [ 
  \frac{1}{r} \, \frac{d f(r)}{d r} + 
  \frac{l(l+1)}{r^2} \right ] \, .
\end{equation}
Using the near horizon approximation, we have 
\begin{equation}
\label{eqn:lmax}
    l^2<\frac{r_{h}^2}{\gamma^2} \quad \mbox{leading to} \quad 
    l_{\rm max}^2\sim\frac{r_h^2}{\gamma^2} \, .
\end{equation}
Thus, the total number of relevant modes are:
\begin{equation}
\label{eqn:NoOfModes}
    N(\gamma)=\sum_{l = 0}^{l_{\rm max}}(2l+1)\sim 
    l_{\rm max}^2=\frac{r_h^2}{\gamma^2}
\end{equation}
As mentioned above, our analysis rests upon the fact that 
the effective 1-D potential has a maximum that traps the high angular momentum modes close to the horizon. In the rest of this work, using two different approaches, we evaluate the length scale at which the back-reaction of the quantum modes on the background metric becomes significant~\cite{Casher}. We refer to the two approaches as 
\emph{Statistical mechanical} (approach 1) and \emph{Quantum Field Theory} (approach 2). 

In approach 1, we obtain the critical length by evaluating the back-reaction of the metric fluctuations and several thermodynamical quantities. In approach 2, we calculate the effect of a shockwave (due to an infalling particle) on the outgoing Hawking particles. The two approaches differ in terms of how the critical length is evaluated. Also, the back-reaction of the fluctuations and the critical length at which the Hawking particles change state vary. 

\section{Approach 1: Statistical Mechanics}
\label{sec:IV}

This section evaluates the physical quantities from statistical mechanics and shows that the quantum scalar field causes an {fluctuation} in the horizon radius. We show that the thermal modes of the field back-react on the metric contributing to the energy of the black-hole. The statistical fluctuations lead to the smearing of the horizon.  We then calculate the invariant distance at which the back reaction due to the field on the metric becomes significant. We do this by including the effects of the statistical fluctuation in the wave equation \eqref{eqn:SchLikeEq}.

Using $N(\gamma)$ obtained in Eq. \eqref{eqn:NoOfModes}, we can calculate the average energy contributed by these modes, which is given by~\cite{1985-tHooft-NPB}:
\begin{equation}
\label{eqn:EnergySM}
    \left \langle E \right \rangle\sim 
    k_B \, T \,  N(\gamma)=\frac{\hbar \kappa}{2\pi}\left(\frac{r_h}{\gamma}\right)^2
\end{equation}
Similarly, the statistical entropy of these thermal modes is~\cite{1985-tHooft-NPB}:
\begin{equation}
\label{eqn:EntropySM}
    S\sim k_B \, N(\gamma) \,  = \, k_B\left(\frac{r_h}{\gamma}\right)^2
\end{equation}
Note that as we move closer to the horizon ($\gamma \to 0$), the average energy and the scalar field's statistical entropy diverge. The fluctuations in the energy of these modes are given by:  
\begin{equation}
\label{eqn:DeltaEnergySM}
    \Delta(E)\sim T\sqrt{k_B\frac{\partial \left \langle E \right \rangle }{\partial T}}\sim \left( \frac{\hbar \kappa}{2\pi} \right)\sqrt{N(\gamma)}= \left( \frac{\hbar \kappa}{2\pi} \right)\frac{r_h}{\gamma}
\end{equation}
 The first part of the above equation is a result from statistical mechanics. If the partition function is $Z\sim\sum{\exp(-\beta E)}$, then average energy is given by $\langle E \rangle =\frac{\sum{E\exp{(-\beta E)}}}{Z}= -\frac{\partial ln Z}{\partial \beta}$ and the average of energy squared is given by $\langle E^2 \rangle =\frac{\sum{E^2\exp{(-\beta E)}}}{Z}= \frac{\partial^2 Z}{\partial \beta^2}\frac{1}{Z}$. If we differentiate the average energy expression again with respect to $\beta$, we get $\frac{\partial \langle E \rangle}{\partial \beta}=\frac{1}{Z^2}\left(\frac{\partial Z}{\partial \beta}\right)^2-\frac{1}{Z}\frac{\partial^2 Z}{\partial \beta^2}=\langle E \rangle^2-\langle E^2 \rangle=-(\Delta E)^2$. Converting the $\beta$ derivative to temperature (T) derivative gives the first part of Eq 27. The subsequent parts of Eq. \eqref{eqn:DeltaEnergySM} are substitutions of different quantities from equations \eqref{eqn:NoOfModes}, \eqref{eqn:EnergySM}, \eqref{def:HawkingTemp}. We want to make a couple of remarks regarding the above result: First, the above result is generic for any spherically symmetric space-time. {Second, like energy and entropy, the energy fluctuations also diverge close to the horizon. The fluctuation in energy leads to a fluctuation in the location of the black-hole horizon.}

In the rest of this section, we use the above expression to obtain the {fluctuation} in the horizon and calculate the length-scale at which the back-reaction becomes relevant. We then apply the results specifically to SAdS and SdS cases.

\subsection{Horizon fluctuations}

Let us consider a point $r_0$ outside the horizon such that,
\begin{equation}
\label{eqn:deltadelat_delta_rh}
    \delta\left(r_{0}\right)=r_{0}-r_{h} > 0 \quad \mbox{and} 
\quad \Delta \delta\left(r_{0}\right)= \Delta r_{h}(r_{0}) \, .
\end{equation}
The {fluctuation} in the black-hole energy leads to the {fluctuation} in the distance from the horizon ($\Delta \delta\left(r_{0}\right)$). (Note that, throughout this subsection, we deal with the absolute values of $\Delta \delta$ and $\Delta r_h$ and hence have dropped the $-ve$ sign in the right hand equation in \eqref{eqn:deltadelat_delta_rh}.) Since $\delta(r_0)$ depends
on the energy of the field modes, we have
\begin{equation}
\label{def:Deltadeltar0}
    \Delta \delta\left(r_{0}\right)= \Delta r_{h}(r_{0})=\frac{\partial (\Delta r_h)}{\partial (\Delta E)}\Delta E
\end{equation}
In the case of Schwarzschild black-hole, $\Delta E \sim\Delta M$ 
(where $M$ is the mass of the black-hole), we have ${\partial (\Delta r_h)}/{\partial (\Delta E)}\sim2$. 
[The quantum scalar field modes contribute to the total energy leading to a change in the energy. For an asymptotic observer, the system's energy content without the scalar field modes is the black-hole mass $ M $. Including the scalar field modes, the system's total energy for the observer at infinity is $M + \mbox{Field's energy content}$. As a consequence of the Birkhoff theorem, this is the new ADM mass of the black-hole space-time. Hence, we can interpret the fluctuation in the energy content of the field as the fluctuation in the ADM mass of the black-hole. Hence, the relation $\Delta E \sim \Delta M$.]  

For $r_{0}$ to be certainly outside the horizon (including the contribution from the field modes), the following condition must be satisfied~\cite{Casher}:
\begin{equation}
    \Delta \delta\left(r_{0}\right)<\delta\left(r_{0}\right) \, .
\end{equation}
To be certain that $ \delta\left(r_{0}\right)$ is positive  and observer independent, it necessary to relate it to the invariant distance. Using \eqref{eqn:GammaDeltaRel1}, we have:
\begin{equation}
\label{eq:deltagamma}
    \delta=\gamma^2\kappa/2
\end{equation}
Substituting Eqs.~\eqref{eq:deltagamma} and \eqref{eqn:DeltaEnergySM} in Eq.~\eqref{def:Deltadeltar0}, 
we get:
\begin{equation}
    \gamma^3>\hbar r_h \, \frac{\partial (\Delta r_h)}{\partial (\Delta E)}
\end{equation}
Therefore, the minimum length up to which the fluctuations smear the horizon is:
\begin{equation}
\label{eqn:HorizonUncertainty}
    \gamma_{_{\rm lim}}=\left[(\hbar r_h) \frac{\partial (\Delta r_h)}{\partial (\Delta E)}\right]^{1/3} \, .
\end{equation}
A closed-form expression can be obtained only for the Schwarzschild black-hole since $\Delta r_h$ and $\Delta E$ are related by a simple relation. For SAdS or SdS black-hole, $\Delta r_h$ is related to $\Delta E$ via transcendental functions and can only be obtained (perturbatively) for a narrow range of parameters. To overcome this, we now use the effective metric approach to evaluate the critical length.

\subsection{Back-reaction on the Metric and Critical Length Scale}

The above analysis provides a minimum length up to which there is a large uncertainty in defining the location of the horizon. The semi-classical Einstein equation describes the effect of the quantum fields on the background space-time~\cite{1984-Birrell.Davies-Book}: 
\begin{equation}
\label{eq:SemiClasEinsEq}
G_{\mu\nu} = 8 \pi \langle T_{\mu\nu} \rangle
\end{equation}
where the source $\langle T_{\mu\nu} \rangle$ is the 
quantum expectation value of the matter field stress tensor operator. Note that the expectation value is only defined after suitable regularization and renormalization~\cite{1993-Kuo.Ford-PRD}. The procedure is well-defined away from the Planck energy~\cite{1993-Kuo.Ford-PRD}. The quantum expectation of the matter field will change the original background metric \eqref{eqn:SSmetric} to:
\begin{equation}
\label{eqn:SSmetricmod}
 d s^{2}=-\widetilde{f}(r) d t^{2}+\frac{d r^{2}}{\widetilde{f}(r)}+r^{2} d \Omega^{2} \, ,   
\end{equation}
where
\begin{equation}
\label{eqn:ftilde}
    \widetilde{f}(r)=f(r)+h(r) \, , 
\end{equation}
and $h(r)$ is the change in the background metric due to the quantum modes. We want to make a couple of remarks regarding the above-modified line-element: First, we have made a specific gauge choice of the perturbed metric. This choice of gauge is to keep the calculations tractable.
It has also been shown that about $90\%$ of Hawking radiation is in $s$-waves~\cite{1976-Page-PRD,1978-Sachez-PRD}. Second, if the back-reaction on the metric is small, $h(r)$ will be small compared to $f(r)$. Naively, this implies that if $h(r) \sim f(r)$, then the back-reaction on the background is significant. In the rest of this sub-section, we quantify this precisely by obtaining an effective 1-D equation for the quantum field $\Phi$ propagating in the above-modified line-element \eqref{eqn:SSmetricmod}.

Substituting the above modified metric in the  scalar field equation \eqref{eq:KGequation} and decomposing the modes into spherical harmonics, we obtain the following effective 1-D differential equation:
\begin{equation}
\label{eqn:Perturbed_Wave_Eq}
    A(r) \partial _{r_{*}}^2 \tilde{u}_{l}(r, \omega) 
    + \left [\omega^2 + C + \left ( \frac{B}{2A} \right ) \frac{\partial A}{\partial r_{*}}-\frac{B^2}{4A}-\frac{1}{2}\frac{\partial B}{\partial r_{*}} \right] \tilde{u}_{l}(r, \omega) =0
\end{equation}
where 
\begin{subequations}
\label{eqn:ABC}
\begin{eqnarray}
\label{eqn:perturbed_Wave_eq_Coeff_A}
   A &=& \left( 1+\frac{h(r)}{f(r)}\right)^2  \\
  \label{eqn:perturbed_Wave_eq_Coeff_b}
    B &=& \left( 1+\frac{h(r)}{f(r)}\right)
    \left(\frac{d h(r)}{d r} 
    -\frac{h(r)}{f(r)}\frac{d f(r)}{d r}\right) \\
    \label{eqn:perturbed_Wave_eq_Coeff_C}
C&=& -\left(f(r)+h(r) \right) 
\left( \frac{1}{r}
\frac{d}{dr}  \left[f(r)+h(r)\right] + \frac{l(l+1)}{r^2} \right )
\end{eqnarray}
\end{subequations}
Note that the above expression is exact. Let us now compare the \eqref{eqn:Perturbed_Wave_Eq} with  Eq.~\eqref{eqn:SchLikeEq}. As mentioned above, in the limit  ${h(r)}/{f(r)} \ll 1$, the two equations are identical, and the effect of the fluctuation on the wave equation can be treated as negligible. This is also evident if we compare the coefficient of $\partial^{2}_{r*}\tilde{u}_{l}(r, \omega)$ in both the equations. This implies that if the following 
condition:
\begin{equation}
\label{eqn: GeneralEqtoCalcGammaCritgen}
\frac{2 h(r)}{f(r)} \sim 1
\end{equation} 
is satisfied, the back-reaction of the quantum field on the metric will be significant. 

We want to make the following two comments about the result: 
First, we have assumed that the metric perturbation $h(r)$ is time-independent.
However, the back-reaction of the field modes on the background metric is generally
time-dependent. In Appendix \eqref{sec:DynamicMetricPerturbation}, we have discussed the more general case in which the perturbation of the metric is time-dependent, i.e. $h(r,t)$. The time-average of $A(r,t)$ is identical to $A(r)$ in Eq. \eqref{eqn:perturbed_Wave_eq_Coeff_A}. Hence, the analysis in this section carries through for time-averaged metric perturbations. Second, this expression is valid for any spherically symmetric space-time. 

\begin{figure}[!htb]
\begin{subfigure}{.5\textwidth}
  \includegraphics[width=1\linewidth]{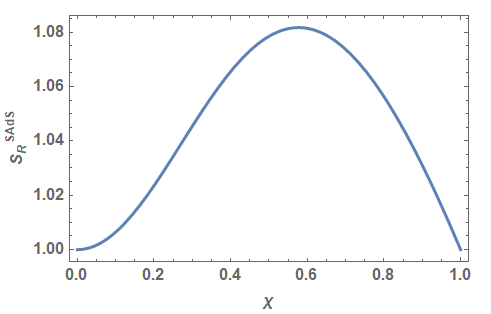}
  \caption{SAdS }
  \label{fig:entropyratioSAdS}
\end{subfigure}%
\begin{subfigure}{.5\textwidth}
  \includegraphics[width=1\linewidth]{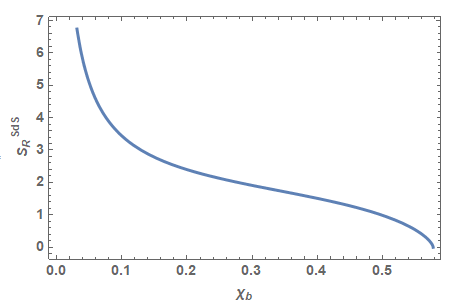}
  \caption{SdS}
  \label{fig:entropyratioSdS}
\end{subfigure}
\caption{Plot of ratio of Entropies with $\chi$ for SAdS and SdS.}
\label{fig:entropyratio}
\end{figure}

In the case of 4-dimensional Schwarzschild, Reissner-Nordstr\"om, SdS and SAdS black-holes, the perturbation of the quantum modes on the metric takes 
the following simple form:
\begin{equation}
    h(r)= \frac{-2\Delta E}{r}
\end{equation}
where $\Delta E$ is given by \eqref{eqn:DeltaEnergySM}. 
Substituting the above expression in \eqref{eqn: GeneralEqtoCalcGammaCritgen} and using \eqref{eqn:DeltaEnergySM}, we have:
\begin{equation}
\label{eqn: GeneralEqtoCalcGammaCrit}
4\left(\frac{\hbar\kappa}{2\pi rf}\right)\left(\frac{r_h}{\gamma}\right)\sim1
\end{equation} 
The assumption of a free scalar field (causing no backreaction on the metric) on a classical background will no longer hold when this condition is satisfied. This condition will give a length-scale in which the original Hawking's derivation can not be trusted. In the rest of this section, we obtain the critical distance, by evaluating the above expression \eqref{eqn: GeneralEqtoCalcGammaCrit}, for three different cases : 
\begin{enumerate}
    \item {\bf Schwarzschild}: In this case, Eq. \eqref{eqn: GeneralEqtoCalcGammaCrit} translates to:
\begin{equation}
\label{eqn:SchGammaCritSM}
    {\gamma^S_{crit}}\sim\left[\frac{8M\hbar}{\pi}\right]^{1/3}
\end{equation}
where $M$ is the mass of the black-hole. The above expression matches with the results of Casher et al. \cite{Casher}. 
{As mentioned earlier, for Schwarzschild,  Eq.~\eqref{eqn:HorizonUncertainty} can be evaluated, and it matches with the above result \eqref{eqn:SchGammaCritSM}.}

\item {\bf SAdS}: In this case, \eqref{eqn: GeneralEqtoCalcGammaCrit} translates to:
\begin{equation}
\label{eqn:SMGammaCritSAdS}
    {\gamma^{(SAdS)}_{crit}} \sim\left[\frac{4}{\pi}\left(\frac{\hbar r_h}{3\left(\frac{r_h}{L}\right)^2+1}\right)\right]^{1/3}=\left[\frac{4L}{\pi}\left(\frac{\hbar \chi}{3\chi^2+1}\right)\right]^{1/3}
\end{equation}
As expected, in the limit $L\rightarrow\infty$, the critical length for the SAdS black-hole reduces to the Schwarzschild result. Since, for SAdS, we can not explicitly obtain a closed-form expression \eqref{eqn:HorizonUncertainty}, we can not compare with the above expression. However, note that both expressions have the same power of $\gamma$ and do point the similarity of both the expressions. 

\item {\bf SdS:} As mentioned earlier, SdS admits two (event and cosmological) horizons. We study the backreaction effect near the event horizon only. 
In Sec. \eqref{sec:Limitations}, we discuss the results for the cosmological horizon. 
In this case, the condition \eqref{eqn: GeneralEqtoCalcGammaCrit} translates to:
\begin{equation}
\label{eqn:dSResultSM}
    \gamma_{crit}\sim \left(\frac{4\hbar L}{\pi}\right)^{1/3}\left(\frac{\kappa_{eff}}{|\kappa_b|}\right)^{1/3}\left(\frac{\chi_b}{1-3\chi_b^2}\right)^{1/3} \, ,
\end{equation}
where,
\begin{equation}
    \chi_b~\in~\left(0,\frac{1}{\sqrt{3}}\right) \, , \,  \quad\quad
    \frac{\kappa_{eff}}{|\kappa_b|}=\frac{\left(2+\frac{\chi_b}{\chi_c}\right)}{\left(4+\frac{\chi_b}{\chi_c}+\frac{\chi_c}{\chi_b}\right)} \, .
\end{equation}
Using \eqref{eqn:chi_dS}, we have:
\begin{equation}
    \chi_c=\sqrt{1-\frac{3}{4}\chi_b^2}-\frac{\chi_b}{2}
\end{equation}
$\chi_b\rightarrow\frac{1}{\sqrt{3}}$ corresponds to the 
the Nariai limit~\cite{Shankaranarayanan:2003ya}. The Nariai limit corresponds to the degenerate horizon and the near horizon assumption being Rindler \eqref{eqn:nearHorizonMetric} {is not valid}. From Eq.~\eqref{eqn:dSResultSM} it is evident that $\gamma_{crit}$ diverges in this limit. In Sec. \eqref{sec:Limitations}, we show that our analysis breaks down much before the Nariai limit. 
\end{enumerate}
Note that in obtaining the critical length for SAdS and SdS, we have not made any assumption about the value of $\chi$. To understand the difference between SAdS and Schwarzschild, we now compute the thermodynamical quantities in the two cases for the same $M$. [While we provide expressions for SAdS, we plot for both SAdS and SdS.] To do that we first write the ratio of $\gamma_{\rm crit}$'s:
    \begin{equation}
\gamma_R=\frac{\rho^{\rm SAdS}_{\rm crit}}{\rho^{\rm S}_{\rm crit}}=\left[\frac{1}{(1+\chi^2)(1+3\chi^2)}\right]^{1/3}
    \end{equation}
Using Eq. \eqref{eqn:EntropySM}, the dimensionless entropy at the critical length scale is:
\begin{equation}
    S^{\rm SAdS}_R=\frac{S^{\rm SAdS}}{S^S} =\left[\frac{(3\chi^2+1)}{(1+\chi^2)^2}\right]^{2/3}=\frac{N^{\rm SAdS}}{N^S}
\end{equation}
Fig. \ref{fig:entropyratio} is the plot of $ S^{\rm SAdS}_R 
(S^{\rm SdS}_R )$ as a function of $\chi$. It is interesting to note that the entropy for the SAdS black-hole is the same as the Schwarzschild black-hole in the obvious case of $\chi\rightarrow 0$ but also in the case $\chi\rightarrow 1$ when $M$ and $L$ are similar. It is important to note that the entropy has a maximum close to $\chi \sim 0.6$. 

Similarly, using Eqs. \eqref{eqn:EnergySM} and \eqref{eqn:DeltaEnergySM}, we obtain dimensionless energy and fluctuations in the energy at the critical invariant distance:
\begin{eqnarray}
\left \langle E\right\rangle^{SAdS}_R &=& \frac{\left \langle E\right\rangle ^{SAdS}}{\left \langle E\right\rangle^S}=\frac{\kappa^{SAdS}N^{SAdS}}{\kappa^S N^S}=\left[\frac{(3\chi^2+1)^5}{(1+\chi^2)}\right]^{1/3} \\
 (\Delta E)^{SAdS}_R &=& \frac{(\Delta E)^{SAdS}}{(\Delta E)^S}=\frac{\kappa^{SAdS}}{\kappa^S }\left(\frac{N^{SAdS}}{N^S}\right)^{1/2}=\left[(3\chi^2+1)^4(1+\chi^2)\right]^{1/3}
\end{eqnarray}

\begin{figure}[!htb]
\begin{subfigure}{.5\textwidth}
  \includegraphics[width=1\linewidth]{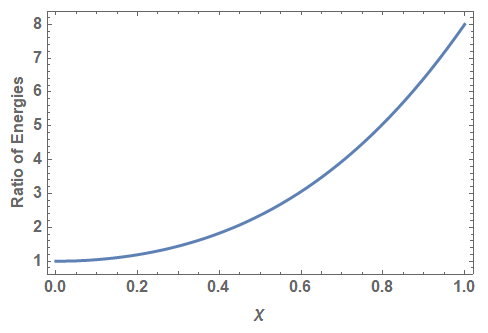}
  \caption{SAdS }
  \label{fig:EnergyRatioSAdS}
\end{subfigure}%
\begin{subfigure}{.5\textwidth}
  \includegraphics[width=1\linewidth]{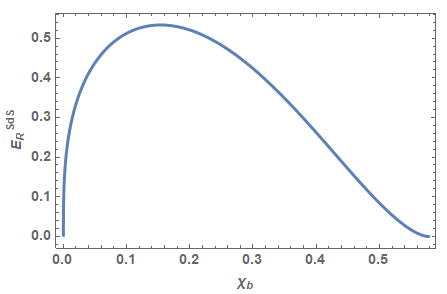}
  \caption{SdS}
  \label{fig:EnergyRatioSdS}
\end{subfigure}
\caption{Plot of ratio of energies with $\chi$ for SAdS and SdS}
\label{fig:EnergyRatio}
\end{figure}

\begin{figure}[!htb]
\begin{subfigure}{.5\textwidth}
  \includegraphics[width=1\linewidth]{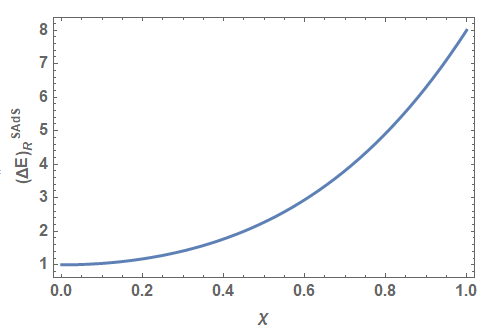}
  \caption{SAdS }
  \label{fig:deltaEnergyRatioSAdS}
\end{subfigure}%
\begin{subfigure}{.5\textwidth}
  \includegraphics[width=1\linewidth]{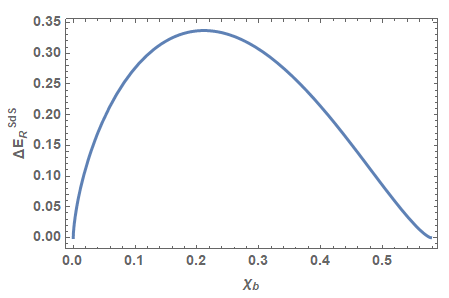}
  \caption{SdS}
  \label{fig:deltaEnergyRatioSdS}
\end{subfigure}
\caption{Plot of ratio of fluctuations in Energy with $\chi$ for SAdS and SdS}
\label{fig:deltaEnergyRatio}
\end{figure}

From Fig. \eqref{fig:EnergyRatio} and Fig. \eqref{fig:deltaEnergyRatio} , we see that the energy and fluctuation in the energy of SAdS increase monotonically with $\chi$. In the case of SdS, the energy and fluctuation in energy each has one maximum. In Sec. \eqref{sec:Limitations}, we discuss the implications of these for our results.

While we were able to obtain a critical length scale using the fluctuations of thermal modes' energy, one of the drawbacks is that we treated $h(r)$ in Eq.\eqref{eqn:SSmetricmod} as a classical perturbation rather than a quantum fluctuation. We will overcome this in Approach 2, which we discuss in the next section. 

\section{Approach 2: Quantum Field Theory}
\label{sec:V}

This section discusses the second approach to obtain a critical length scale at which the gravitational interactions become significant. We study the interaction of the thermal particles in the black-hole atmosphere, with a point-like massless particle falling radially into the black-hole ($s$-wave). (The reason for this choice is because it has been shown numerically that about $90\%$ of Hawking radiation is in $s$-waves~\cite{1976-Page-PRD,1978-Sachez-PRD}.) A massless point-like particle propagating in the Minkowski space-time produces a shock-wave with which the thermal particles of the atmosphere interact~\cite{Shockwave, Aichelburg:1970dh}. 

As discussed in Sec. \eqref{sec:II}, we will use the shock-wave analysis in the Minkowski space-time to study the near horizon properties in any non-degenerate spherically symmetric black-hole space-time. This is because the near-horizon geometry can be approximated as a Rindler, which has a one-to-one mapping with the Minkowski space-time~\cite{Rindler_Ref1,Culetu:2002td}. We show that the shock-wave interaction with the thermal particles becomes significant
as the infalling particle reaches a critical invariant distance from the horizon. We assume that the infalling particle ($s-$wave), with momentum $p$, to move along the $v-$direction ($u=u_0=$ constant). 
The line-element for the space-time with the shock-wave is~\cite{Shockwave,Casher}:
\begin{equation}
\label{eqn:ShockwaveMetric}
    d s^{2}=-d u\left[d v+2 p \ln \left(\frac{\tilde{x}^{2}}{r_h^{2}}\right) \delta\left(u-u_{0}\right) d u\right]+d x^{2}+d y^{2}
\end{equation}
where $\widetilde{x}^2=x^2+y^2$ and $r_h$ is the horizon radius. We would like to make a few remarks regarding the above line-element:
First, the presence of $r_h$ in the line-element is to provide a length-scale to compare with $\widetilde{x}$. Thus, in principle, $r_h$ can be replaced by mass or energy of the black-hole as these also provide a length-scale of the system. Second, the above line element is different from the Minkowski line-element~\eqref{eqn:MinkoskiUVmetric} just due to a shift in the $v-$direction:
\begin{equation}
    v\rightarrow v+2 p \ln \left(\frac{\tilde{x}^{2}}{r_h^{2}}\right)\Theta (u-u_0)
\end{equation}
Hence, the shock-wave line-element remains the same as Minkowski everywhere except in the $u=u_0$ hyperplane. See Appendix \eqref{sec:App} for details on the shock-wave interaction with thermal particles.
\newline

Our interest is to obtain the critical length where the interaction between the shock-wave and thermal particles is significant and can not be neglected. One way to calculate the critical length is to calculate the probability of the thermal particles to be in the same state after interaction~\cite{Casher} with the shockwave. Thus, 
the probability will be close to unity if the interaction is weak; however, it will be negligible if it is strong. To go about calculating this probability, we describe the thermal particles by a wavepacket ($g(k)$) with a small spread [that depends on $l$, cf. \eqref{eqn:lmax}]:
\begin{equation}
  \phi(x)=\langle x  |\phi\rangle=N \int dk \, g(k) \, e^{i k(x-T)}
\end{equation}
$g(k)$ can, in principle, be any function which fixes the momentum width of the wave packet. Using $|\left\langle\phi | \phi\right\rangle|^{2}=1$, we fix $N$ to be $\frac{1}{\sqrt{2\pi}}$. The effect of the shockwave on the thermal particles can be described by the shift in $v$ coordinate given by:
\[
2p \ln\left( \frac{\widetilde{x}^2}{M^2} \right) \sim p\, ,
\]
where $p$ is the momentum of the incoming particle (along the $v$-direction) and is a constant of motion (For details, see Appendix \eqref{sec:App}.) Note that the $\log$ term will not contribute much in the final expression for the probability. Hence, the wave packet after crossing the
shockwave would be:
%
\begin{equation}
 \phi'(x)=\langle x   |\phi'\rangle \simeq N \int dk\, g(k)\, e^{i k\left(x-T+p\right)} \, ,
\end{equation}
Using the fact that the momentum of a thermal particle near the horizon is $1/\gamma_0$ (Appendix \eqref{sec:AppB}) before it interacts with the shockwave, we choose $g(k)$ to be a Gaussian wave-packet:
\begin{equation}
    |g(k)|^2=\frac{1}{2\pi (\Delta k)^2}e^{-\frac{-(k-\bar{k})^2}{2(\Delta k)^2}}
\end{equation}
where 
\begin{equation}
\label{eqn:widthPeak_of_wavepacket}
    \Delta k=\bar{k}=\frac{1}{\gamma_{0}}
\end{equation}
We have chosen $\Delta k$ to be of the same order as $\bar{k}$.
Thus, the probability of the thermal particle to be in the same state even after interacting with the shockwave is given by:
\begin{equation}
    P=\left|\left\langle\phi^{\prime} | \phi\right\rangle\right|^{2}\sim
    \exp\left(-\frac{4\lambda^2}{\kappa^2\gamma_0^4}\right)
\end{equation}
where $\lambda$ is related to the momentum of the infalling massless particle (See Eq.~\eqref{eqn:plambdaRelationApp}):
\begin{equation}
      \lambda\sim \frac{p\,\kappa\,\gamma_0}{2} \quad \mbox{and} \quad 
      \gamma_0\sim \frac{r_h}{l} \, .
\end{equation}
Note that $\gamma_0$ is the maximum invariant distance that any thermal particle can reach. Probability of the entire thermal atmosphere to not change state is:
\begin{equation}
\label{eqn:ProbAtmosphere}
    P_{\rm tot}\sim \exp\left( - \frac{4 \lambda^2}{\kappa^2} 
    \sum_{l } \frac{(2l+1)}{\gamma_{0(l)}^4} \right) \sim  \exp\left(-\frac{4\lambda^2}{\kappa^2\gamma^4}N(\gamma)\right)
\end{equation}
In obtaining the second expression, we approximate the summation by treating $ \gamma $ to be a free parameter. Eq. \eqref{eqn:ProbAtmosphere} is a vital result regarding which we would mention the following points: First, the result we have obtained is different from the one obtained by Casher et al. \cite{Casher}. We choose a particular $g(k)$ to get the exact result for the probabilities instead of making approximations. Our results match with that in Ref.~\cite{Casher} in the limit when the exponential is approximated to the first two terms in the series expansion. Second, We have obtained the probability for a general spherically symmetric metric.

From the above expression \eqref{eqn:ProbAtmosphere}, we can obtain the critical length at which the interactions become strong. When the interaction is weak, most thermal particles will remain in the same state. This corresponds to:
\begin{equation}
\label{eqn:ProbablityArgumentForWeakInteraction}
\frac{4\lambda^2}{\kappa^2\gamma^4}N(\gamma)<<1 \, . 
\end{equation}
This implies that if the following condition:
\begin{equation}
    \frac{4\lambda^2}{\kappa^2\gamma^4}N(\gamma)\sim 1
\end{equation}
is satisfied, the interaction between the shockwave and the thermal particles become significant. This tells us that up to the following invariant distance from the event horizon:
\begin{equation}
\label{eq:App2Gamma}
    \gamma\sim\left(\frac{2\lambda r_h}{\kappa}\right)^{1/3}\sim \left(\frac{r_h^2\,p}{l}\right)^{1/3}
\end{equation}
the interactions cannot be neglected. Note that \eqref{eq:App2Gamma} is related to $p$ (constant of motion). Physically, for a black-hole with horizon $r_h$, $p$ can not take arbitrarily small values. More specifically, the de Broglie wavelength of the incoming particle can not be larger than the event horizon. If the de Broglie wavelength is larger than the horizon radius, there will not be any interaction. Thus, a black-hole with event horizon $r_h$ sets the minimum de Broglie wavelength of the incoming particle. This leads to the following condition:
\begin{equation}
    \frac{h}{p}<r_h \, \implies \,  
    p > \left(\frac{2\pi\hbar}{r_h}\right)
\end{equation}
Substituting the minimum possible value of $p$ 
in Eq. \eqref{eq:App2Gamma}, ignoring prefactor of $\pi/l$, we have:
\begin{equation}
\label{eqn:approach2Result}
    \gamma_{\rm crit}\sim(2\hbar r_h)^{1/3}
\end{equation}
This is an important result regarding which we would like to stress the following points: First, the critical distance we have obtained above is for any spherically symmetric space-time~\eqref{eqn:SSmetric} with the event horizon at $r_h$. In approach 1, we could obtain a closed-form expression only for a class of spherically symmetric black-hole space-times. However, in approach 2, we do not any make any such assumption. Second, recently, such a feature has emerged in describing the fluid-gravity correspondence leading to the quantization of horizon area~\cite{2017-Cropp.etal-PRD,2019-Bhattacharya.Shanki-Arxiv}.

To compare and contrast the two approaches, in the rest of this section, we evaluate the critical distance \eqref{eqn:approach2Result} for the three black-hole space-times: 
\begin{enumerate}
    \item {\bf Schwarzschild}: In this case, Eq. \eqref{eqn:approach2Result} reduces to:
\begin{equation}
\label{eqn:SresultApp2}
    {\gamma^S_{\rm crit}}\sim\left(4\hbar M\right)^{1/3}
\end{equation}
where $M$ is the mass of the black-hole. This matches with the expression \eqref{eqn:SchGammaCritSM} (besides the overall constant). Also, the above expression matches with the corresponding results of Casher et al.~\cite{Casher}. 

\item {\bf SAdS}: In this case, Eq.~\eqref{eqn:approach2Result} translates to:
\begin{equation}
\label{eqn:SAdSresultApp2}
\gamma_{\rm crit}^{\rm SAdS} \sim (2\hbar L)^{1/3}\chi^{1/3}
\end{equation}

\item {\bf SdS:} Near the event horizon, Eq.~ \eqref{eqn:approach2Result} reduces to:
 \begin{equation}
 \label{eqn:SdSresultApp2a}
    \gamma_{\rm crit}^{\rm SdS} \sim (2\hbar r_b)^{1/3}\sim (2\hbar L)^{1/3}\chi_b^{1/3}
\end{equation}
\end{enumerate}
To get a numerical intuition of the critical length scales, let us calculate the numerical value of $\gamma_{crit}$ using Eq.~\eqref{eqn:approach2Result} for a supermassive Schwarzschild blackhole with a billion solar mass. In SI units, Eq.~\eqref{eqn:approach2Result} can be rewritten as:
\[
\gamma_{crit}=(2\hbar r_h)^{1/3}=(2l_{\rm pl}^2 r_h)^{1/3} \, .
\]
where $l_{pl}=\sqrt{\hbar G/c^3}$. For $M=10^9 M_{\odot}$, we have
\[
r_{h}=\frac{2 G M}{c^2}\sim 2.95\, \times \,10^{12}\, m \, .
\]
Substituting $l_{\rm pl} \sim 10^{-35} m$ and $r_h$ in $\gamma_{crit}$, we get
$\gamma_{crit} \sim 1.541\,\times \,10^{-19}\,m $. The value is extremely small (order $10^{-19}\, m$) because $\gamma_{crit}=(2\hbar r_h)^{1/3}=(2l_{pl}^2 r_h)^{1/3}$ in terms of the Schwarzschild radius $r_h$, is
suppressed by the Planck length ($l_{pl}$). Thus, we see that
for a supermassive black-hole of a billion solar masses, the critical length scale (width of the shell) at which the back-reactions become significant is of the order of $10^{-19}\, m$. 

Before we proceed with the limitations of the analysis, we want to
highlight the following points: First, in obtaining the critical length for SAdS and SdS, we have not assumed the value of $\chi$. Second, these results are consistent with the results obtained by Sorkin using a different procedure~\cite{Sorkin:1995ej}.  Third, both the approaches can not be trusted beyond $\gamma_{crit}$. Around $\gamma_{crit}$, the gravitational interactions become strong, and the shock-wave analysis~\cite{Shockwave} ceases to hold. Lastly, the critical length for both the approaches match for Schwarzschild, they differ for SAdS \eqref{eqn:SMGammaCritSAdS} and SdS \eqref{eqn:dSResultSM}. In the next section, we identify the cause of the deviation between the two approaches for SAdS and SdS and point to the limitations of the analysis.

\section{Limitations of the analysis}
\label{sec:Limitations}

\begin{figure}
\begin{subfigure}{.5\textwidth}
  \centering
  \includegraphics[width=1\linewidth]{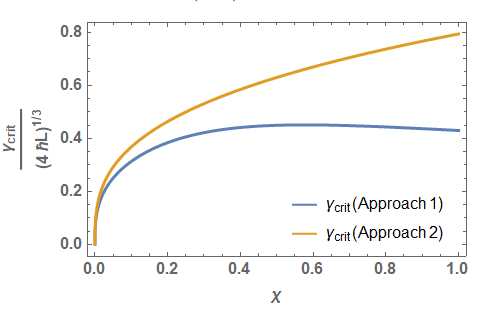}
  \caption{SAdS}
  \label{fig:GammaCrit2Cases_SAdS}
\end{subfigure}%
~
\begin{subfigure}{.5\textwidth}
  \centering
  \includegraphics[width=1\linewidth]{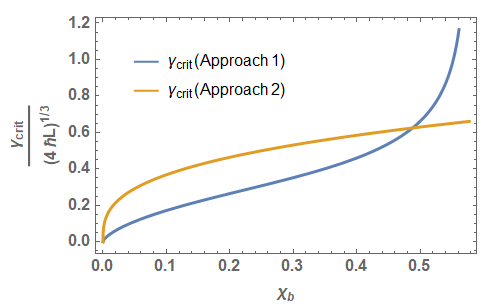}
  \caption{SdS}
  \label{fig:GammaCrit2Cases_SdS}
\end{subfigure}
\caption{Plot of the critical length $\gamma_{\rm crit}/(4 \hbar L)^{1/3}$ as a function of $\chi$ ($\chi_b)$ in the two approaches for SAdS and SdS.}
\label{fig:GammaCrit2Cases}
\end{figure}

Using two different approaches, for general spherically symmetric black-hole space-times, we obtained a critical scale at which the quantum scalar field's backreaction to the metric becomes significant. While the critical length in the two approaches matches for Schwarzschild black-hole, the functional dependence of the critical length differs in the two approaches for SAdS and SdS black-holes. 
The plots in Fig. \eqref{fig:GammaCrit2Cases} highlight the difference between the critical length ($\gamma_{\rm crit}$) as a function of $\chi$ in the two approaches for SAdS and SdS.  

This is a crucial observation regarding which we want to highlight the following points: First, for SAdS and SdS, the two approaches seem to have different profiles. In the small $\chi$ $(\chi_b)$ limit, they agree. Second, in the case of SAdS, while approach 1 predicts finite critical length for all $\chi$, approach 2 predicts a growing critical length for large values of $\chi$. Note that this is the limit ($\chi \to 1$)
considered in AdS/CFT correspondence where the large black holes in
AdS are dual to the high-temperature phase of the dual field theory~\cite{Hubeny}. Third, in the case of SdS, while approach 2 predicts finite critical length for all possible values of $\chi$, approach 1 predicts a diverging critical length for $\chi_b \to 1/\sqrt{3}$. 

\begin{figure}[!htb]
    \centering
    \includegraphics[scale=0.9]{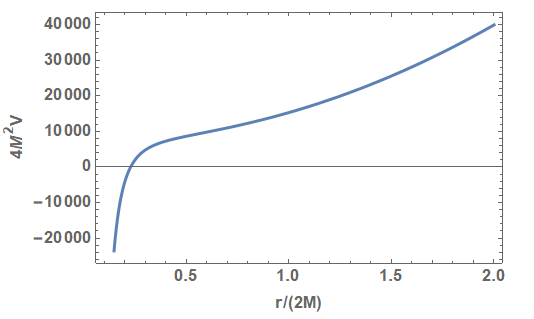}
    \caption{No Peak in SAdS Potential near horizon for $\frac{M}{L}=4$, ($l=10$)}
    \label{fig:NoPeakPotential}
\end{figure}

This leads to the important question: Why is there a discrepancy between the two approaches for SAdS and SdS? As mentioned in Sec. \eqref{sec:III}, our analysis rests on the fact that there is a peak (positive barrier) in the potential profile $V(r)$ that localizes the high angular momentum modes close to the horizon. 
However, 
this assumption may not be true for all values of $\chi$. In
Fig.~\eqref{fig:NoPeakPotential}, we have plotted the potential $(4 M^2 V)$ for SAdS for specific values of $M/L = 4$ and $l = 10$. As we can see, the peak in the potential is not present, and the potential increases monotonically. Compare this with the plots in Fig. \eqref{fig:PotentialSAdS_SdS}, where the potential has a peak at a finite value from the horizon. Specifically, we observe the absence of a peak near the horizon in the potential for ${M}/{L}\gtrsim 1.5$ and any $l<100$. Thus, as we increase the value of $M/L$ (corresponding to increasing values of $\chi$), we observe that the peak in the potential slowly vanishes.

\begin{figure}[!htb]
    \centering
    \includegraphics[scale=0.9]{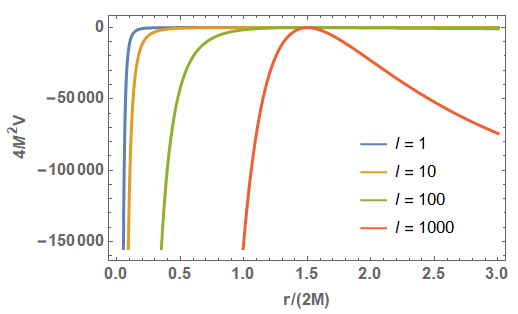}
    \caption{SdS Potential (is neagative) near horizon for $\frac{M}{L}=0.193$, for different values of l}
    \label{fig:dSNegativePotential}
\end{figure}

Fig.~\eqref{fig:dSNegativePotential} is the plot of the potential $(4 M^2 V)$ for the event horizon of SdS for  $M/L = 0.193$. The plot highlights that the potential is negative for any value of $l$. 
Hence, Eq.~\eqref{eqn:OmegaGreaterPotential} will not provide any finite value of $l_{\rm max}$. {Beyond a range of M/L values for SdS, the number of relevant modes is not finite.}

This leads to the following: First, there is no upper bound on the angular momentum implying that all the angular momentum modes have to be considered. This means that $\gamma \to 0$ and 
the semi-classical analysis is bound to breakdown. Second, for Approach 1, from Eqs. (\ref{eqn:EnergySM}, \ref{eqn:DeltaEnergySM}), we have:
\begin{equation}
\label{eqn:DeltaEbyE}
\frac{\Delta E}{\langle E\rangle}\sim \frac{1}{\sqrt{N}}\sim 0 \, ,
\end{equation}
implying that the backreaction on the metric is negligible. 
Third, for approach 2, this limit implies that the condition \eqref{eqn:ProbablityArgumentForWeakInteraction} is never true.
Physically, this means that gravitational interactions are strong everywhere. When the gravitational interactions are strong everywhere, the shock-wave metric \eqref{eqn:ShockwaveMetric} is not a valid metric in this limit. Hence, our analysis will not hold in this limit.

As mentioned earlier, both the approaches require a finite value of $l_{\rm max}$ (cf. \eqref{eqn:lmax}) which is true only if the effective 1-D potential $V(r)$ \eqref{Eqn:RGPotential} has a maximum near the horizon. We have shown that the two approaches give similar 
critical length if the conditions (\ref{eqn:OmegaGreaterPotential}, \ref{eqn:lmax}) are satisfied.

\section{Conclusions and Discussions}

Hawking's original derivation ignores the radiation's interaction with matter and requires the incoming vacuum modes with frequencies far above the Planck scale. One way to address both these assumptions is not to make the horizon a fixed surface, but a dynamical one that changes due to the backreaction of the quantum modes~\cite{Casher,Sorkin:1995ej}. Such an approach is also consistent within the framework of effective field theory~\cite{1994-Donoghue-PRD,2004-Burgess-LRR}. 

In this work, we aimed to ask until which length scale the quantum modes' backreaction can be ignored. We identified the length scale where the assumption of a free field on a classical background to explain Hawking radiation is not valid, and the quantum modes smear the horizon. While there have been attempts in the literature, our analysis is different for two specific reasons: First, we have obtained the length scale for an arbitrary 4-dimensional spherically symmetric space-time. 
We also showed that for a supermassive Schwarzschild black-hole of a billion solar masses, the critical length scale (width of the shell) at which the back-reactions become significant is of the order of $10^{-19}\, m$. 
Second, we have identified the limitations of the approach in the asymptotic non-flat space-times. 

We used two different approaches to evaluate the critical length. 
In the first approach, we used the statistical mechanical properties of the perturbations and calculated the change in the mass of the black-hole due to these fluctuations. In the second approach, we calculated the effect of a shock-wave (due to the thermal particles) on the outgoing Hawking particle. While the two approaches use two different features of the fluctuations, both the approaches assume a finite value of $l_{\rm max}$. We showed that as long as this condition is satisfied, both the approaches provide similar results. The results of approach 2 are consistent with Ref.~\cite{Sorkin:1995ej}.

We showed that this condition is explicitly violated for SAdS and SdS for a finite value of $M/L $. In SdS, as $M/L $ is increased, the potential is negative for any value of $l$. Hence, our results do not apply to the cosmological horizon and in the Nariai limit. In the case of SAdS, we showed that as $M/L$ is increased (corresponding to increasing values of $\chi$), the peak in the potential vanishes. The difference in the Free energy of  $\mathrm{AdS}_{d+1}$ with the $\mathrm{BH}$ solution ($F_{\rm bh}$) and 
free energy of $\mathrm{AdS}_{d+1}$  without black-hole ($F_{\rm th}$) is given by~\cite{1982-Hawking.Page-CMP}
\begin{equation}
F_{\rm bh}-F_{\rm th}=\frac{r_{h}^{d -1} V_{S^{d - 1}}}{d r_{h}^{2}+(d-2) L^{2}}\left(L^{2}-r_{h}^{2}\right)
\end{equation}
where $V_{S^{d - 1}}$ is the volume of the $(d - 1)$-sphere.
{The above expression implies that $r_h = L$ is the critical size.} For $r_h < L$, the thermal AdS is less energetic solution while the opposite is true in the limit $r_h > L$. Our results imply that critical distance where the back-reaction becomes strong can be computed in the limit of $M/L < 1$ ($r_h/L < 1$). 

Counter-intuitively, our results are more accurate in the late stages of the evaporation process. As black-holes lose mass due to evaporation, $M/L$ decrease, and both the approaches give similar results. However, for the Planck-size black-holes, General Relativity itself needs to be modified to include higher-order Riemann tensor, Ricci tensor and scalar terms \eqref{eq:EFT}. In that scenario, the background metric line-element \eqref{eqn:SSmetric} itself may be modified. It is important to note that our calculations do not use any of the physical inputs of the Firewall proposal~\cite{2013-Almheiri.etal-JHEP}, however, our results may have implications for the Firewall proposal. This is currently under investigation.

One of the reasons for choosing a massless scalar field is that the gravitational perturbations about the background metric have the same form~\cite{1973-Misner.etal-Gravitation,2004-Das.Shanki-CQG}. For the modified gravity theories, we need to obtain the corresponding equation to study the back-reaction. This is currently under investigation. 

Our results bring attention to the following interesting questions: Are there modifications to the Hawking temperature, especially in the limit $M/L \sim 1$ (for SAdS)? How does the Hawking rate change for these space-times? How is the Page time \cite{1993-Page-PRL} affected in the limit $M/L \sim 1$ (for SAdS)? We hope to return to study some of these questions soon.

\section*{Acknowledgements} \label{sec:acknowledgements}
The authors thank M. Bojowald, T. Dray, Ted Jacobson, Sudipta Mukherji, T. Padmanabhan, P. Ramadevi, A. Virmani, and U. Yajnik for discussions. We are grateful to Mahendra Mali for his help during the initial stages of this work. We thank Saurya Das and S. Mahesh Chandran for helpful comments on an earlier draft.
This work is part of the Undergraduate project of MS. The work of SS is supported by the MATRICS SERB grant (MTR/2019/000077) and Homi Bhabha Fellowship.

\appendix
\section{Shockwave Analysis}
\label{sec:App}

Here we discuss the effect of a point-like massless particle propagating in the Minkowski background. 
The Minkowski line-element is given by:
\begin{equation}
    ds^2=-dT^2+dz^2+dx^2+dy^2 \, ,
\end{equation}
where 
\begin{equation}
\label{eqn:uvDef}
    u=T+z\quad\quad\quad\quad v=T-z \, .
\end{equation}
This leads to the following form of the Minkowski metric:
\begin{equation}
\label{eqn:MinkoskiUVmetric}
    ds^2=-dudv+dx^2+dy^2
\end{equation}
The above line-element can describe the region near the event horizon of 4-D spherically symmetric black-hole space-time 
described by the line-element \eqref{eqn:SSmetric}. To see this, 
we define the following:
\begin{equation}
\label{eqn:uv_kappaGamma}
    u=\gamma e^{\kappa t}\quad\quad\quad v=-\gamma e^{-\kappa t}
\end{equation}
where $\kappa$ is the surface gravity defined in \eqref{eqn:kappaGeneral} and $\gamma$ is given by \eqref{eqn:gammaDef}. Under this transformation, the above 
line-element \eqref{eqn:MinkoskiUVmetric} leads to:
\begin{equation}
ds^2= -\kappa^2\gamma^2dt^2+d\gamma^2+dx^2+dy^2
\end{equation}
Note that the above line-element corresponds to the near-horizon line-element of a spherically symmetric black-hole obtained in 
\eqref{eqn:nearHorizonMetric} in the $x-t$ plane. The $2-$Sphere part of the metric can also be mimicked (See \cite{Rindler_Ref1,Culetu:2002td}). Dray and t'Hooft~\cite{Shockwave} showed that a massless particle moving in Minkowski space produces a \emph{shockwave}. Due to the mapping described above, the analysis of Dray and t'Hooft~\cite{Shockwave} can be extended to any horizon.  In the 
rest of this Appendix, we give the key results which are essential for our calculations. The gravitational field of a massless particle in Minkowski space moving along the $v$ direction ($u=u_0$), with momentum $p$, is described by the line element \cite{Casher}:
\begin{equation}
\label{eqn:ShockwaveMetric1}
    d s^{2}=-d u\left[d v+2 p \ln \left(\frac{\tilde{x}^{2}}{r_h^{2}}\right) \delta\left(u-u_{0}\right) d u\right]+d x^{2}+d y^{2}
\end{equation}
with $\widetilde{x}^2=x^2+y^2$ and $r_h$ is the horizon radius. $r_h$ just provides a scale (in the system concerned) to compare $\widetilde{x}$ with. $r_h$ in \eqref{eqn:ShockwaveMetric1} could, in principle, be replaced by the mass or the energy of the black-hole as these also provide a scale in the system.. Note that the above line element is different from the Minkowski case only by a shift in the $v$ direction given by:
\begin{equation}
\label{eqn:vShift}
    v\rightarrow v+c \, \Theta (u-u_0)
\end{equation}
with 
\begin{equation}
c=2 p \ln \left(\frac{\tilde{x}^{2}}{r_h^{2}}\right)
\end{equation}
Therefore the space remains same as Minkowski everywhere except in the hyperplane $u=u_0$. The effect of such a shockwave on other particles can be studied using the Hamilton Jacobi formalism. The Hamilton Jacobi equation is given by ($S$ is action):
\begin{equation}
    -4\left(\frac{\partial S}{\partial u}\right)\left(\frac{\partial S}{\partial v}\right)+8 p \ln \left(\frac{\tilde{x}^{2}}{r_h^{2}}\right) \delta\left(u-u_{0}\right)\left(\frac{\partial S}{\partial v}\right)^2+\left(\frac{\partial S}{\partial x}\right)^2+\left(\frac{\partial S}{\partial y}\right)^2=0
\end{equation}
At $u\neq u_0$ (absence of \emph{shockwave}), the solution is given:
\begin{equation}
\label{eqn:S_wo_Shockwave}
    S_0=\hbar\left[k_xx+k_yy+k_vv+\left(\frac{k_x^2+k_y^2}{4k_v}\right)u\right]
\end{equation}
Here the $k$'s are the wave vectors of the thermal particles before encountering the shockwave. In the presence of shockwave, it is natural to use \eqref{eqn:vShift} in \eqref{eqn:S_wo_Shockwave} to get an action upto the leading correction, which gives:
\begin{equation}
    S=S_{0}+2 p^{v} \ln \left(\frac{\tilde{x}^{2}}{r_h^{2}}\right) \hbar k_{v} \theta\left(u-u_{0}\right)+ \mbox{Subleading terms}
\end{equation}
This action can then be used to calculate shifts (refractions) in the $x$, $y$ directions as:
\begin{equation}
k_{x}(u)=\frac{\partial S}{\partial x}=k_{x}+\frac{4 p^{v}}{\tilde{x}^{2}} x k_{v} \theta\left(u-u_{0}\right)+ \mbox{Subleading terms}
\end{equation}
\begin{equation}
k_{y}(u)=\frac{\partial S}{\partial x}=k_{y}+\frac{4 p^{v}}{\tilde{x}^{2}} y k_{v} \theta\left(u-u_{0}\right)+ \mbox{Subleading terms}
\end{equation}

\section{Calculating the momentum of the Hawking Particles before interaction with the  Shockwave}
\label{sec:AppB}
Here, we approximate the momentum of the thermal particles before they interact with the shockwave. In line-element \eqref{eqn:nearHorizonMetric}, the massless particles follow the geodesics (with $g_{\mu\nu} V_0^\mu V_{0}^{\nu}=0$):
\begin{equation}
    X^\mu=X^\mu _0+\alpha V_0^\mu \, .
\end{equation}
Using a suitable rotation, boost and translation,
the above geodesic of the massless particle takes the following form~\cite{Casher}:
\begin{equation}
\label{eqn:MasslessParticleTrajectory}
    X^\mu(\alpha)=\left(T=-\gamma_0+\alpha,\quad y=0, \quad x=-\gamma_0+\alpha, \quad z=\rho_0 \right)\quad\quad 0<\alpha<2\gamma_0 \, ,
\end{equation}
where $\gamma_0\sim \frac{r_h}{l}$ is the maximum invariant distance that any thermal particle can reach. While the above expression is general and can be applied to any physical situation, our interest is near-horizon. We have used Eq.~\eqref{eqn:lmax} to relate $\gamma_0$ with the orbital angular momentum number $\ell$ and the critical distance. 

As in Ref. \cite{Casher}, we assume that the ingoing particle (that produces the shockwave, which is also massless) has momentum $p$ only along the 
$v$-direction. In other words, it travels at a fixed $u$-coordinate.  

Now we calculate the $\gamma$ at which a thermal particle crosses the shockwave. At that $\gamma$, we will try to estimate the momentum of the thermal particle.
Using the thermal particle's trajectory \eqref{eqn:MasslessParticleTrajectory} for a fixed $u$ equal to that of the ingoing particle, we get (using \eqref{eqn:uvDef} and \eqref{eqn:uv_kappaGamma}):
\begin{equation}
    \gamma^2_{cross}=2\gamma_0u-u^2
\end{equation}
We make the assumption that most thermal particles that can reach the maximum $\gamma=\gamma_0$, cross the shockwave at $\gamma_{cross}=\gamma_0$ itself. I.e. most of the thermal particles that can reach $\gamma_0$ have already reached $\gamma=\gamma_0$ before the shockwave arrives. Using this approximation and the above equation, we get:
\begin{equation}
    \gamma_{cross}\sim\gamma_0\sim u
\end{equation}
Now we calculate the Schwarzschild energies ($\lambda$, negative of the time component of the four momentum covariant vector) of any particle is our spacetime:
\begin{equation}
    \lambda=-p_t=\kappa^2\gamma^2p^t
\end{equation}
which after a short calculation in terms of $u$ and $v$ gives:
\begin{equation}
\label{eqn:SchEnergyUVGeneral}
    \lambda=-p_t=\kappa(vp_v-up_u)
\end{equation}
For the infalling particle (s-wave):
\begin{equation}
    p^v=p;\quad p^u=0; \quad p_v=0; \quad p_u=-\frac{p}{2}
\end{equation}
Thus we get a relation between the momentum of the particle and its Schwarzschild energy as:
\begin{equation}
\label{eqn:plambdaRelationApp}
    p=\frac{2\lambda}{\kappa u}\sim \frac{2\lambda}{\kappa \gamma_0}
\end{equation}
For thermal particles we approximate the Schwarzschild energy to be equal to the $k_BT\sim\hbar\kappa$. Also for the thermal particles, near the horizon the $u$ and $v$ direction very nearly coincide (horizon is a null surface), we have $\hbar k_u=\hbar k_v$. Using this in \eqref{eqn:SchEnergyUVGeneral}, we get:
\begin{equation}
\label{eqn:WavepacketWidth}
    k_u\sim \frac{1}{\gamma_0}
\end{equation}
Equation \eqref{eqn:WavepacketWidth} gives an estimate of the momentum of a thermal particle before it crosses the horizon. To fix the width and the peak of the wave packet of the thermal particle before interaction with the shockwave we use \eqref{eqn:WavepacketWidth} as an estimate. 

\section{Effect of the time-dependent metric perturbation on the critical length scale}
\label{sec:DynamicMetricPerturbation}

In Sec. \eqref{sec:IV}, we obtained the critical length at which the back-reaction of the quantum field on the metric will be significant assuming that the metric perturbation is time-independent. In this Appendix, we relax the condition and assume that the metric perturbation is time-dependent:
\begin{equation}
\label{eq:backreaction}
  	\tilde{f}(r,t)=f(r)+h(r,t)
  \end{equation}
We can use the following ansatz for the generic time-dependent field configuration (not necessarily harmonic):
    \begin{equation}
   	\phi_{l m}=\frac{u_{l}(r, t)}{r} Y_{l m}(\theta, \phi) 
   	\end{equation}
 Substituting the above ansatz in the wave equation \eqref{eq:KGequation}, we have:
   	\begin{equation}
   	\label{eqn:AppWaveEq}
   		-(\partial_{t}\tilde{f}\partial_{t}u_l+\partial_{t}^2u_l)+\partial_{\tilde{r}^*}^2u_l-\tilde{f}\left(\frac{l(l+1)}{r^2}+\frac{1}{r}\partial_{r}\tilde{f}\right)u_l=0
   	\end{equation}
 where $\tilde{r}^*$ is:
  \begin{equation}
  	\tilde{r}^*=\int\frac{dr}{\tilde{f}}
  \end{equation}
Substituting the above form of the metric \eqref{eq:backreaction} in the above wave-equation, we have 
\begin{eqnarray}
\label{eqn:Unperturbed_Wave_Eq}
\mbox{Unperturbed Eq.} & & ~~~ -\partial_{t}^2u_l+\partial_{r^*}^2u_l-f\left(\frac{l(l+1)}{r^2}+\frac{1}{r}\partial_{r}f\right)u_l=0 \\
  \label{eqn:perturbed_Wave_Eq}
\mbox{Perturbed Eq.} & & ~~~
-\partial_{t}^2\tilde{u}_{l}(r, t) + A(r,t) \partial _{r^{*}}^2 \tilde{u}_{l}(r, t) \\
 & &  + \left [ C + \left ( \frac{B}{2A} \right ) \frac{\partial A}{\partial r^{*}}-\frac{B^2}{4A}-\frac{1}{2}\frac{\partial B}{\partial r^{*}} +\frac{1}{2}\frac{\partial^2 h}{\partial t^2}+\frac{1}{4}\left(\frac{\partial h}{\partial t}\right)^2\right] \tilde{u}_{l}(r, t) =0 \nonumber
\end{eqnarray}
 where $A$, $B$, and $C$, defined in Eq \eqref{eqn:ABC}, are time-dependent. 
 Note that the above equation is similar to \eqref{eqn:Perturbed_Wave_Eq} except for 
\[
\frac{1}{2}\frac{\partial^2 h}{\partial t^2}u_l+\frac{1}{4}\left(\frac{\partial h}{\partial t}\right)^2u_l
\]
terms. Our interest is to study the effect of the fluctuations over a long period. Mathematically, this corresponds to replacing the above expression with time averages (which would be only functions of $r$). Thus the time-dependence leads to a change in the potential of the perturbed wave equation; however, it does not change the coefficients of $\partial_{r^*}^2 \tilde{u}_l$ terms, and the time average of $A(r,t)$ corresponds to $A(r)$ in \eqref{eqn:ABC}. Hence, the analysis in Sec. \eqref{sec:IV} follows.

\nocite{*}

\bibliography{apssamp}

\end{document}